\documentclass[preprint,groupedaddress,superscriptaddress,amsmath,amssymb]{revtex4-1}[10pts]
\usepackage{float}
\usepackage{graphicx}
\usepackage{dcolumn}
\usepackage{bm}
\usepackage{color}
\usepackage{comment}
\usepackage{soul}
\usepackage{ulem}
\graphicspath{{figures/}}
\begin{document}

\preprint{AIP/123-QED}

\title[]{A \textbf{multi-GHz} chaotic optoelectronic oscillator based on laser terminal voltage}

\author{C. Y. Chang}
\email{cychang@gatech.edu} 
\affiliation{Georgia Institute of Technology, School of Physics, Atlanta, Georgia 30332-0250 USA}%
\affiliation{%
UMI 2958 Georgia Tech-CNRS, Georgia Tech Lorraine, 2 Rue Marconi F-57070, Metz, France}%

\author{Daeyoung Choi}%
\affiliation{%
UMI 2958 Georgia Tech-CNRS, Georgia Tech Lorraine, 2 Rue Marconi F-57070, Metz, France}%
\affiliation{School of Electrical and Computer Engineering, Georgia Institute of Technology, 
Atlanta, Georgia 30332-0250 USA}%

\author{A. Locquet}
\affiliation{%
UMI 2958 Georgia Tech-CNRS, Georgia Tech Lorraine, 2 Rue Marconi F-57070, Metz, France}%
\affiliation{School of Electrical and Computer Engineering, Georgia Institute of Technology, 
Atlanta, Georgia 30332-0250 USA}%

\author{Michael J. Wishon}
\affiliation{%
UMI 2958 Georgia Tech-CNRS, Georgia Tech Lorraine, 2 Rue Marconi F-57070, Metz, France}%
\affiliation{School of Electrical and Computer Engineering, Georgia Institute of Technology, 
Atlanta, Georgia 30332-0250 USA}%

\author{K. Merghem}

\author{Abderrahim Ramdane}
\affiliation{%
CNRS, Laboratory for Photonics and Nanostructures, Route de Nozay, 91460 Marcoussis, France}%

\author{Fran\c{c}ois Lelarge}
\affiliation{%
CNRS, Laboratory for Photonics and Nanostructures, Route de Nozay, 91460 Marcoussis, France}%
\affiliation{%
III-V Lab, a joint Laboratory of Alcatel Lucent Bell Labs and Thales Research \& Technology 
and CEA-LETI, Route de Nozay, 91460 Marcoussis, France}%

\author{A. Martinez}
\affiliation{%
CNRS, Laboratory for Photonics and Nanostructures, Route de Nozay, 91460 Marcoussis, France}%

\author{D. S. Citrin}
\email{david.citrin@ece.gatech.edu}
\affiliation{%
UMI 2958 Georgia Tech-CNRS, Georgia Tech Lorraine, 2 Rue Marconi F-57070, Metz, France}
\affiliation{School of Electrical and Computer Engineering, 
Georgia Institute of Technology, Atlanta, Georgia 30332-0250 USA}%

\date{\today}
\begin{abstract}
A multi-GHz chaotic optoelectronic oscillator based on an external cavity semiconductor laser (ECL) is 
demonstrated.   Unlike standard optoelectronic oscillators for microwave applications, 
we do not employ the dynamic light output incident on a photodiode 
to generate the microwave signal, but instead generate the microwave 
signal directly by measuring the terminal voltage $V(t)$
of the laser diode of the ECL under constant-current operation, 
thus obviating the photodiode entirely.

\end{abstract}

\maketitle

Microwave signals are commonly used in communications, radar, and medical imaging. 
High-frequency microwave waveforms are conventionally generated in the electrical domain 
using digital electronics and typically involve several stages of multipliers and amplifiers.
This approach, however, may be inefficient at high frequencies.
Another possibility is to generate microwave waveforms in the optical domain, to take advantage 
of the broad bandwidth and low attenuation in the optical system. 
Optical generation of microwave signals enables tremendous flexibility; the optical signal can be converted 
immediately to a microwave electrical signal via a fast photodiode (PD) or can be transmitted over 
low-loss optical fiber systems to be converted downstream to a microwave electrical signal.

Specific implementations of such optoelectronic oscillators (OEOs) can be based on the beating of two 
phase-locked optical waves\cite{yao1,yao2}, through OEOs based on optical injection of a master laser 
into a slave laser\cite{15Nelson} or electro-optic 
modulators\cite{96Yao,10Chembo} have been demonstrated. 
Moreover, on-chip OEO miniaturization and integration is possible\cite{malekiNP}, providing a further
desirable feature of OEOs. 

In addition to the interest in optical microwave generation of periodic\cite{yao1} and shaped-pulse 
\cite{yao2} signals, there has also been 
a stream of work on the generation of broadband chaotic signals through the use of 
electro-optic modulators subjected to feedback\cite{96Celka,02Goedg} of a 
laser diode (LD) subjected to optoelectronic\cite{01JMLiu} or 
optical feedback\cite{92Mork,95Celka} (ECL), or to optical injection\cite{97Simpson,kuwashima3}. 
It is notable that certain applications\cite{13Soriano} of chaotic laser dynamics, 
and in particular chaotic radar\cite{9} and ultrahigh-rate 
random-bit generation \cite{6,7,nianqiang1}, do not intrinsically make use of the chaotic optical
signal, but of an electrical signal detected by one or more PDs.
These comments also apply for the work cited above on the generation of periodic microwave signals: 
a PD separate from the LD is always necessary to create an electrical signal while the optical signal 
itself is often not used directly [unless low-loss optical transmission of the signal is needed prior to 
optical-to-electrical (O/E) conversion]. Clearly, circumventing  
O/E conversion could significantly simplify various applications of such chaotic signals.

\begin{figure}[b]
\centerline{\includegraphics[width=.6\textwidth]{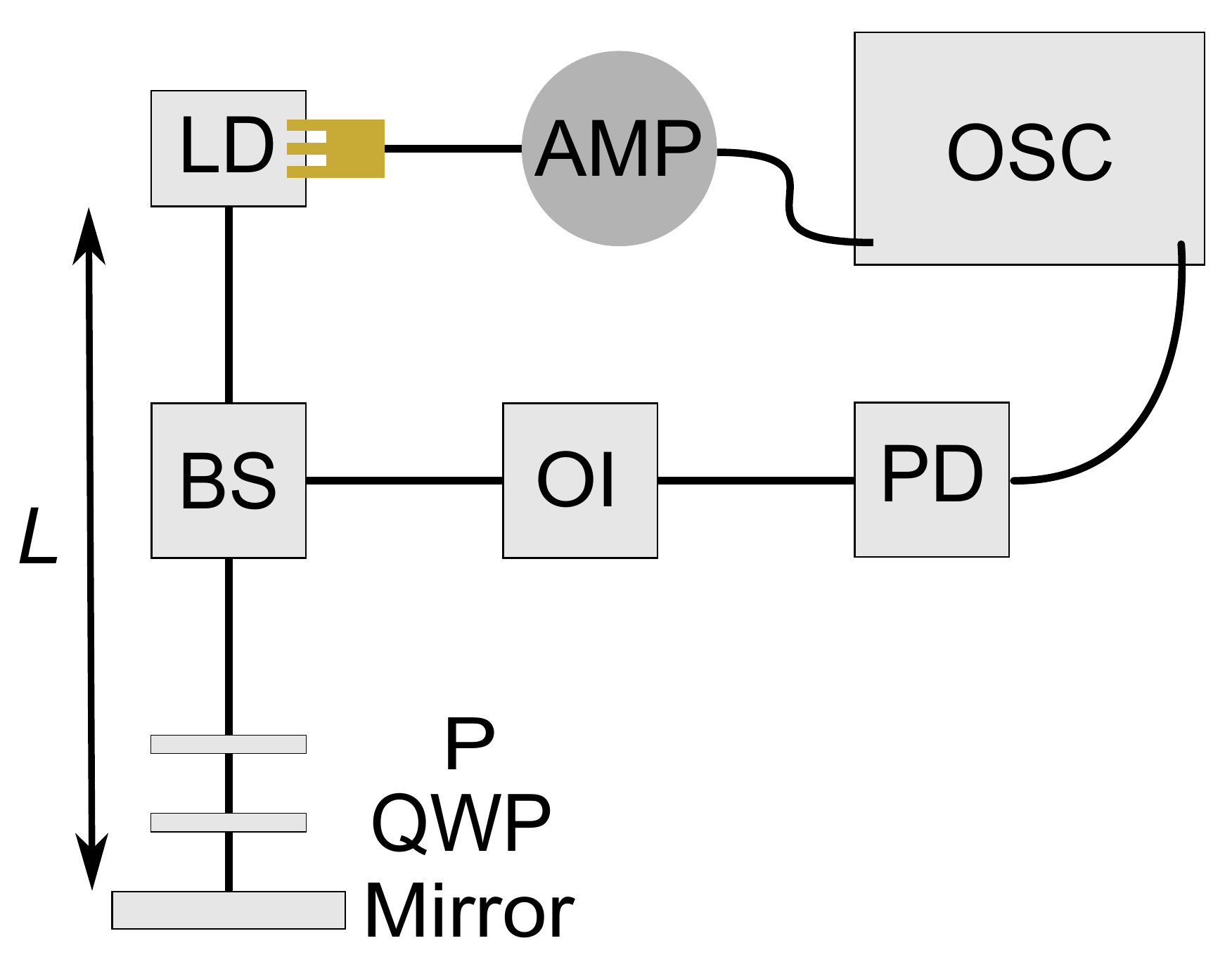}}
\caption{Experimental setup. LD: laser diode, AMP: RF amplifier, OSC: oscilloscope, BS: beam splitter, 
OI: optical isolator, PD: photodiode, P: polarizer, QWP: quarter-wave plate.
}
\label{figure1}
\end{figure}


In this letter, we demonstrate a chaotic multi-GHz OEO based on an ECL 
with \textit{direct} chaotic microwave electrical generation.  In other words, our approach entirely obviates the use of 
a PD by directly monitoring the time-dependent voltage $V(t)$ across the injection
terminals of the laser diode under constant-current $J$ injection.  
We verify that the dynamics exhibited by $V(t)$ is indeed chaotic and of comparable dynamical complexity as that of 
$I(t)$ by means of a largest-Lyapunov-exponent (LLE) analysis. The basis for our observation 
is that for small signals, $V(t)$ is proportional to the inversion $N(t)$ in the gain
medium, as was pointed out in Refs.\ \onlinecite{kazarinov,sahai,Roy}.  
The dynamics of $N(t)$ and $I(t)$, in turn, are closely linked,
as is understood, for example, on the basis of the Lang-Kobayashi equations \cite{lk}. 
Ref. \onlinecite{FischerTomo} recently showed a measurement of the voltage $V(t)$ across 
the LD but focused uniquely on its use, in conjunction with a phase measurement, 
to describe and understand the regime of low-frequency fluctuations.

\begin{figure}[ht]
\vspace{-3mm}
\centerline{\includegraphics[width=0.6\textwidth]{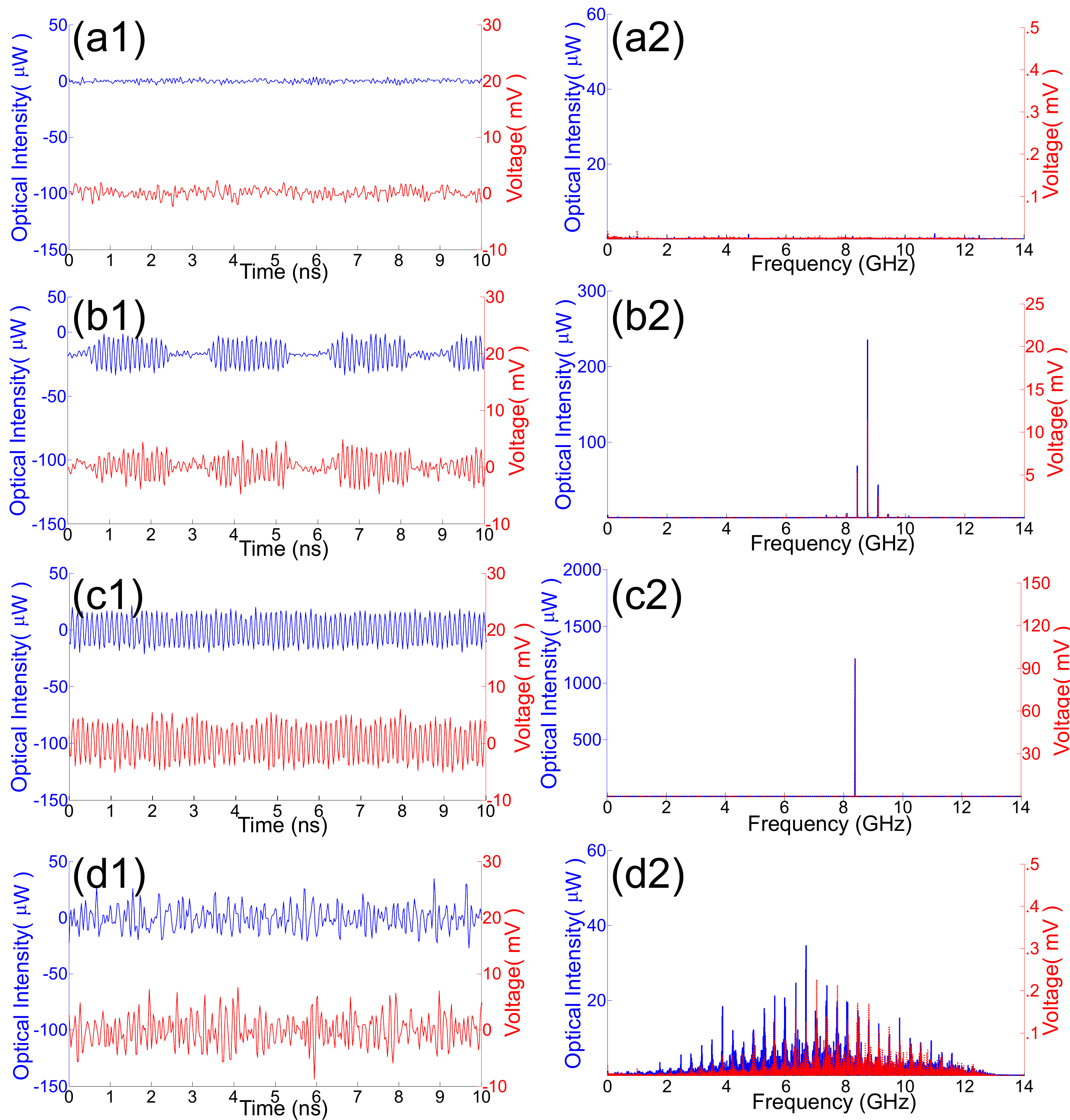}}
\caption{The left panels show the time series of $I(t)$ 
and $V(t)$ in blue and red, respectively, for various $\eta=0.01, 0.19, 0.28$, and $0.77$ from top to bottom. 
The right panels represent the RF spectra of each time series shown on the left. Panels (a1, a2) 
correspond to CW behavior, (b1, b2) to quasi-periodic, (c1,c2) to periodicity, and (d1,d2) to chaotic behavior.
}
\label{figure2}
\end{figure}

The experimental setup is shown in Fig.\ 1. The
single longitudinal-mode edge-emitting InGaAsP multi-quantum-well 
DFB LD emits at 1550 nm with free-running threshold current $J_{th}=29.8$ mA. The heterostructure, containing 7 quantum wells, and the grating are designed and fabricated to achieve a $k$ product of 50 cm$^{-1}$ and the length $l$ of the laser diode is measured to be $0.6$ mm, resulting in a $kl$ value of $3$.
The detailed structure has been described and investigated for feedback tolerance in Ref. \onlinecite{AnthonyAPL}.
The experimental feedback strength $\eta$ is determined by the relative angle 
between the polarizer and the quarter-wave plate (QWP), where $\eta = 1$ 
corresponds to maximum feedback strength $\eta_{max}$ ($\sim$16 \% of the 
optical power that is coupled back onto the collimating lens). The QWP is mounted on a motorized rotational 
stage with a step size of $0.01^{\circ}$.  
During the experiment, a RF probe (Cascade Microtech AE-ACP40-GSG-400) with bandwidth of 40 GHz is used to extract 
$V(t)$ from the LD injection terminals. 
The AC component of $V(t)$ is separated from the DC component with a bias tee (Keysight 11612A), 
and then amplified with an 18-dB amplifier (Newport 1422-LF) with 20 GHz bandwidth. 
The AC components of intensity, $I(t)$ and voltage, $V(t)$ are both 
captured by a real-time oscilloscope (Agilent DSO80804B)
whose cut-off frequency is 12 GHz. The external cavity length $L$ is chosen to be 30, 42, or 70 cm 
corresponding to external cavity round-time $\tau= 2$, 2.80, or 4.67 ns and giving an 
external-cavity free-spectral range of
$f_{\tau} = \tau^{-1}=0.5$, 0.35, or 0.21 GHz, respectively. 

Figure 2 shows $I(t)$ and $V(t)$ with their RF 
spectra for various $\eta$ at  $J = 70.12$ mA and $L = 42$ cm via FFT of the original time series.
Under these conditions, the relaxation-oscillation frequency, $f_{RO}=8.04$ GHz and 
$f_{\tau}=0.35$ GHz. Both $I(t)$ and $V(t)$ are extracted 
simultaneously from the oscilloscope and synchronized precisely by calibrating the time delay 
in the optical versus the electrical path. We observe that the ECL experiences a range of dynamical 
regimes (CW, periodic, quasi-periodic, and chaotic) and can be accessed both from the optical 
intensity and the laser voltage. We also observe, as expected, that $V(t)$ and $I(t)$ are typically highly 
anticorrelated on the timescale of $f_{RO}^{-1}$ as carriers recombine to emit photons thereby reducing $N(t)$.

\begin{figure}[ht]
\centerline{\includegraphics[width=0.6\textwidth]{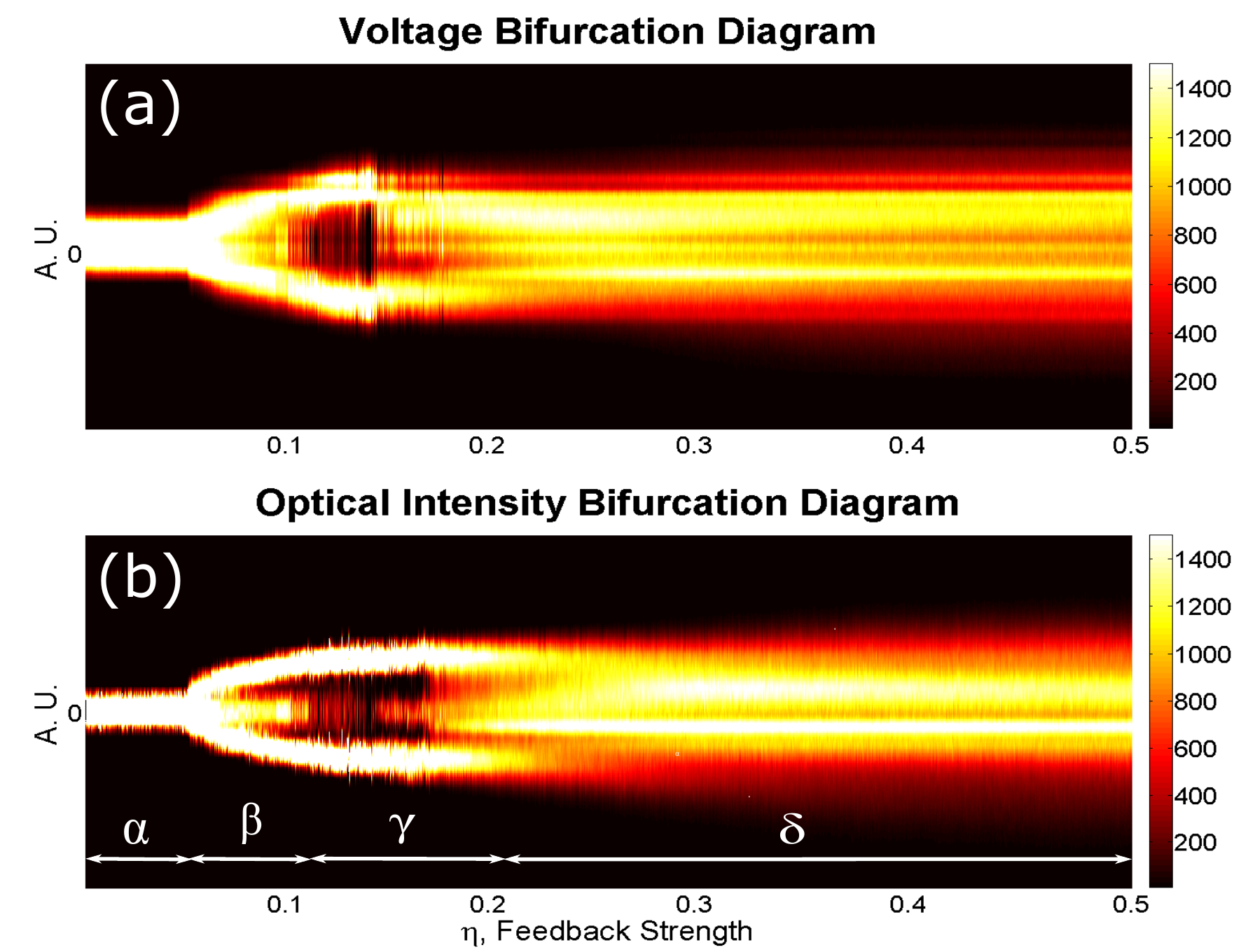}}
\caption{(a) Voltage $V(t)$ and (b) optical intensity $I(t)$ bifurcation diagrams (BDs) for 
$J = 70.12$ mA and $L = 42$ cm as feedback strength $\eta$ is increased.}
\vspace{-5mm}
\label{figure3}
\end{figure}

Figure 3 shows the simultaneously measured bifurcation diagrams (BDs) for $V(t)$ and $I(t)$.
BDs have been obtained recently for the optical $I(t)$ alone\cite{bobbyOE,bobbyPRA}.
Each BD is obtained by plotting the density of local extrema of the corresponding time series as the $\eta$ is 
increased. In both BDs, the density is high in white but low in black. Both Figs. \ 2 and  3  indicate the dynamical regime of the ECL \cite{bobbyPRA}. 

The simultaneous examination of Figs.\ 2 and 3 reveals the broad range of dynamical behaviors of an ECL-based OEO.  In the following, without going into the details of all the 
bifurcations experienced by the ECL, we present the structurally stable dynamical regimes, 
i.e. those that remain qualitatively the same in a large region of values of $\eta$.
When $\eta< 0.06$, corresponding to region $\alpha$ in Fig.\ 3(b),  the ECL 
is in continuous-wave (CW) operation. 
Fig.\ 2(a) further illustrates that the noise level is of similar magnitude on 
$I$ and $V$, though slightly larger on the voltage. 

Region $\beta$ in Fig.\ 3(b) shows quasi-periodic behavior of the 
ECL, which is confirmed by the spectra of $I$ and $V$ that both 
show the presence of two incommensurate frequencies, in the form 
of a central peak at 8.74 GHz, and sidebands offset by 0.35 GHz 
from the central peak. This quasi-periodic behavior is expected 
to occur on the route to chaos for an ECL\cite{92Mork}, where the central peak frequency is close to that of the
relaxation-oscillation frequency $f_{RO}$ and that the side-peak separation is close to $f_\tau$. The undamping of these two frequencies probably results from two successive Hopf bifurcation in  Ref. \onlinecite{92Mork} and Ref. \onlinecite{bobbyPRA}. Periodic dynamics are observed within region $\gamma$, and in Fig. 2(c), where a single peak dominates the RF spectrum. In this case, the peak is at $\sim 8.40$ GHz, reflecting $f_{RO}$ plus $f_\tau$, as already observed in Ref.\ \onlinecite{bobbyPRA} in the case of 
a multi-quantum-well laser. The observation of such periodic windows in 
the route to chaos is consistent with previous reports
Ref. \onlinecite{92Mork} and Ref. \onlinecite{ bobbyPRA} of periodic windows in the quasi-periodic route to chaos in an ECL. Moreover, the observed frequency of 
$\approx 8.46$ GHz in region $\gamma$, reflecting $f_{RO} + f_\tau 
$, is also consistent with our previous conclusion in Ref. \onlinecite{bobbyPRA}.

\begin{figure}[ht]
\centerline{\includegraphics[width=0.7\textwidth]{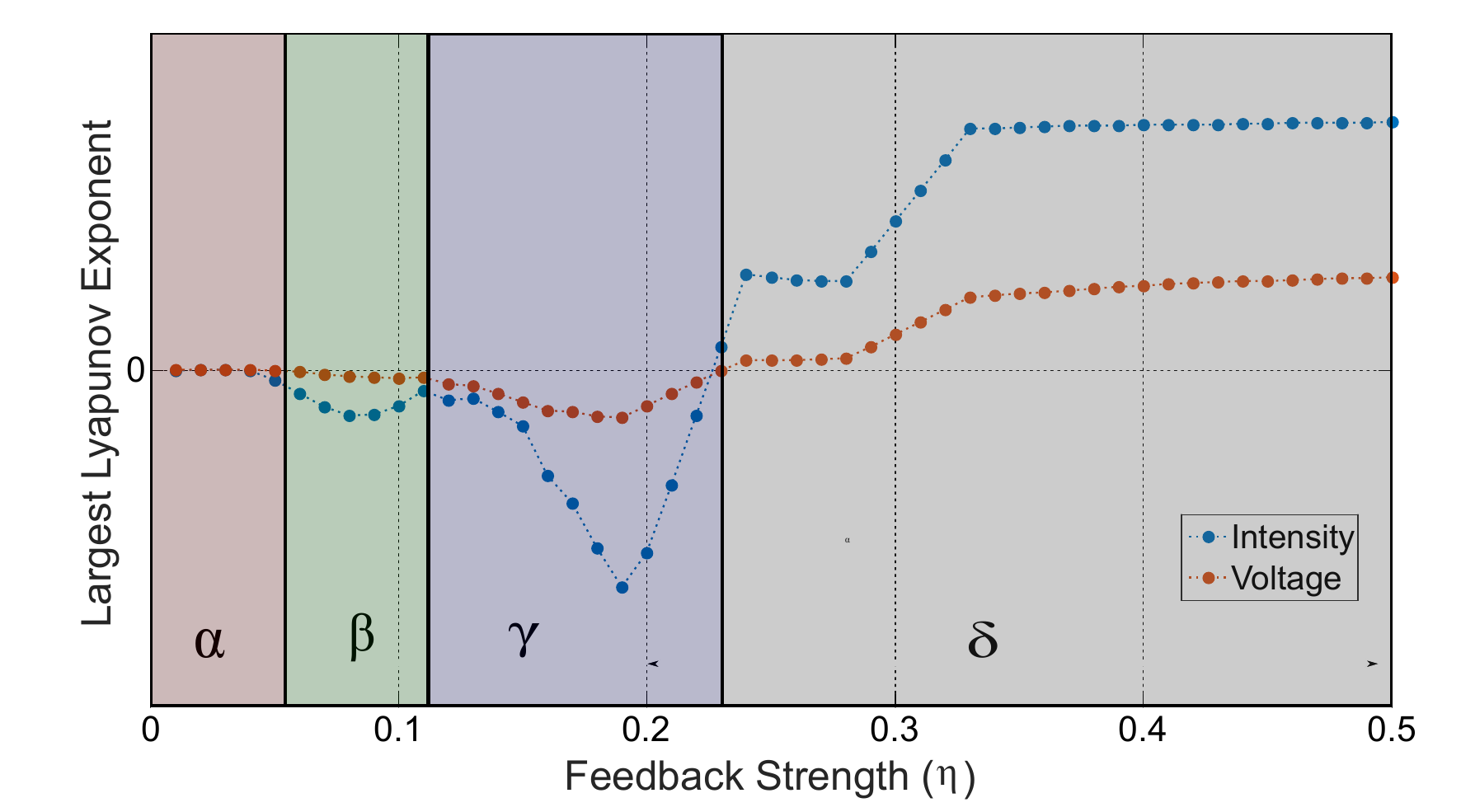}}
\caption{The largest Lyapunov exponent is plotted as a function of increasing feedback strength for both the optical intensity and voltage across the laser diode. The value is negative or close to zero when we are in the CW ($\alpha$), quasi-periodic ($\beta$), and periodic ($\gamma$) regimes. As we enter the coherence collapse regime ($\delta$) the exponent becomes positive indicating chaos.}
\label{figure4}
\vspace{-3mm}
\end{figure}

As we further increase the feedback strength above $\eta>0.23$, we enter region $\delta$ of the BDs (Fig.\ 3), $V(t)$ and $I(t)$ reveal an apparently erratic behavior [Fig.\ 2(d1)] and the corresponding spectra show considerable broadening [Fig.\ 2(d2)], as the 6 dB bandwidth is approximately $4.56$ GHz. The suspected chaotic behavior in region $\delta$ is confirmed by evaluating the (LLE) from the time series of both $I$ and $V$\cite{HeggerLLE, KantzPRA,KantzLLE}. In Fig. 4, the value of LLE is plotted as a function of feedback strength for both optical intensity and voltage. We obtain consistently a (strictly) positive and similar value of the LLE from the analysis of $I$ or $V$ in region $\delta$ ($\eta>0.23$), confirming the chaotic nature of the dynamical 
behavior. Although further work must be 
carried out, these results indicate that both $I(t)$ and $V(t)$ exhibit comparable chaotic characteristics. 
Moreover, we observe an increase of the LLE with $\eta$, indicating tunability of chaos complexity 
in the proposed OEO. Although not shown here, we have further confirmed the robustness of 
these observations by resolving BDs for 
various values of $L$ (30, 42, and 70 cm) and $J$ 
(50, 60, and 70 mA) which enables further tunability of the RF spectrum and chaos complexity. 
In region $\delta$, the bandwidth of $V(t)$ where the ECL is in coherence collapse (well developed chaos) 
extends up to 8GHz.

As ECLs are recognized as being simple and inexpensive sources of high-dimensional 
chaos\cite{FischerPRL}, as is of interest for applications such 
as chaos radar \cite{9} and ultrahigh rate 
random-bit generation \cite{6,7,nianqiang1}, our work demonstrates the ability to generate such 
microwave signals \textit{directly}, and \textit{not} requiring O/E conversion. 
In conclusion, we have demonstrated a multi-GHz OEO that entirely circumvents the need for 
O/E conversion. We have shown that the voltage $V(t)$ across the 
LD allows for the robust generation of periodic, quasi-periodic, as well as chaotic oscillations, 
of various complexity, that have similar characteristics to those obtained from the optical intensity. 

We gratefully acknowledge the financial support of the 2015 Technologies Incubation scholarship from Taiwan Ministry of Education, the Conseil Regional de Lorraine, and the Fonds European de Developement Regional (FEDER).

\end{document}